# Image Authentication Technique in Frequency Domain based on Discrete Fourier Transformation (IATFDDFT).

Nabin Ghoshal[1], J. K. Mandal[2]

[1]*Department of Engineering and Technological Studies, University of Kalyani, Kalyani, Nadia-741235, West Bengal, India, e-mail:- nabin_ghoshal@yahoo.co.in*

[2]*Dept. of Computer Science and Engineering, University of Kalyani, Kalyani, Nadia-741235 West Bengal, India, e-mail:- jkm.cse@gmail.com.*

**Abstract-** **In this paper a novel data embedding technique in frequency domain has been proposed using Discrete Fourier Transform (DFT) for image authentication and secured message transmission based on hiding a large volume of data into gray images. Image authentication is done by embedding message/image in frequency domain by choosing image blocks of size $2 \times 2$, called mask, from the source image in row major order and transform it into the frequency domain using DFT. Three bits of authenticating message/image/message-digest are fabricated within the real parts of each source image byte except first frequency component of each mask. The dimension of authenticating image followed by message digest (MD) and the content of authenticating message/image are also embedded. Inverse DFT (IDFT) is performed on embedded data to transform embedded frequency component to spatial component. In order to keep the quantum value positive and non negative in spatial domain a strong and robust technique is incorporated mainly on the first frequency component and sometimes on other component depends upon situations. The decoding is done by applying the reverse algorithm. Experimental results conform that the proposed algorithm performs better than DCT, QFT and SCDFT schemes.**

*Keywords*- Image Authentication Technique in Frequency Domain based on Discrete Fourier Transformation (IATFDDFT)**,** QFT, DFT, IDFT, DCT and SCDFT.

## I. INTRODUCTION

Steganography is the art of hiding information into picture or other media in such a way that no one apart from the sender and intended recipient even realizes that there is hidden information. Image transmission via the internet has some problem such as information security, copyright protection, Originality etc. Secured communication is possible with the help of encryption technique which is a disordered and confusing message that makes suspicious enough to attack eavesdroppers. Without creating any special attention of attackers steganographic methods [1, 2, 3] overcome the problem by hiding the secrete information behind the source image. Image trafficking across the network is increasing day by day duo to the proliferation of internetworking. Image authentication is needed to prevent unauthorized access in various e-commerce application areas. This security can be achieved by hiding data within the image. Data hiding [4, 5, 6, 7, 10] in the image has become an important technique for image authentication and identification. Therefore, military, medical and quality control images must be protected against attempts to manipulations. Generally digital image authentication schemes mainly falls into two categories-spatial-domain and frequency-domain techniques. So, digital image authentication [12, 13] technique has become a challenging research area focused on battling to prevent the unauthorized or illegal access and sharing.

So many works has been done in spatial-domain for digital image authentication. Among these the most common methods Chandramouli et al. [8] developed a useful method by masking, filtering and transformations of the least significant bit (LSB) on the source image. Dumitrescu et al. [9] construct an algorithm for detecting LSB steganography. Pavan et al. [11] and N. N.

EL-Emam [5] used entropy based technique for detecting the suitable areas in the image where data can be embedded with minimum distortion. Ker [14] and C. Yang [15] presented general structural steganalysis framework for embedding in two LSBs and Multiple LSBs. H.C. Wu [16] and C-H Yang [17] constructed LSB replacement method into the edge areas using pixel value differencing (PVD).

Several works has been done in frequency domain for digital image authentication. In this area most common transformations are the discrete cosine transformation (DCT), quaternion Fourier transformation (QFT), discrete Fourier transformation (DFT), discrete wavelet transformation (DWT), and the discrete Hadamard transformation (DHT). Frequency-domain methods are widely applied than the spatial-domain methods. Here embedding is done in the frequency component of the image pixel in frequency-domain the human visual system is more sensitive to low frequency components than the high frequency component. To avoid severe distortion of the original image the midrange frequencies are best suitable for embedding to obtain a balance between imperceptibility and robustness. I. J. Cox et al. [18] developed an algorithm to inserts watermarks into the frequency components and spread over all the pixels. DCT-based image authentication is developed by N. Ahmidi et al. [19] using just noticeable difference profile [20] to determine maximum amount of watermark signal that can be tolerated at each region in the image without degrading visual quality. P. Bas et al. [21] proposed a color image watermarking scheme using the hypercomplex numbers representation and the quaternion Fourier transformation. Vector watermarking schemes is developed by T. K. Tsui [22] using complex and quaternion Fourier transformation.

The proposed IATFDDFT emphasizes on information and image protection against unauthorized access in frequency domain to achieve a better tradeoff between robustness and perceptibility. This paper aims to exploit embedding process invariant of positive or negative frequency component. This paper used the Discrete Fourier Transform to get frequency component for each pixel value. The Discrete Fourier Transform (DFT) of spatial value f(x , y) for the image of size M x N is defined in equation 1 for frequency domain transformation.

$$F(u, v) = \frac{1}{\sqrt{MN}} \sum_{x=0}^{M-1} \sum_{y=0}^{N-1} f(x, y)\, e^{-j2\pi\left(\frac{ux}{M} + \frac{vy}{N}\right)} \quad (1)$$

where u = 0 to M – 1 and v = 0 to N-1.

Similarly inverse discrete Fourier transform (IDFT) is used to convert frequency component to the spatial-domain value, and is defined in equation 2 for transformation from frequency to spatial-domain.

$$f(x, y) = \frac{1}{\sqrt{MN}} \sum_{u=0}^{M-1} \sum_{v=0}^{N-1} F(u, v)\, e^{j2\pi\left(\frac{ux}{M} + \frac{vy}{N}\right)} \quad (2)$$

where u = 0 to M – 1 and v = 0 to N-1.

This paper presents a technique for image protection by inserting large amount of messages/image along with message digest MD into the source image for image identification and also for secure message transmission. In IATFDDFT using 24 bits color image, 3 bits of secrete data are inserted in each of the red, green and blue components from LSB, so a total 9 authenticating bits can be inserted in each pixel. Insertion is done at even and odd positions in alternate image byte. IATFDDFT embeds large amount of authenticating message/image with a bare minimum change of visual pattern with better security against statistical attacks.

Section II of the paper deals with the proposed technique. Results, comparison and analysis are given in section III. Conclusions are drawn in section IV and references are given in section V.

II. THE TECHNIQUE

IATFDDFT used 24 bit colour image in which each pixel is the composition of red (R), green (G) and blue (B) of each 8-bit image. The proposed IATFDDFT embeds authenticating message/image $AI_{p,q}$ of size 3*(m x n) bits along with 128 bits MD and dimension of authenticating message/image (32 bits) to authenticate the source image $SI_{m,n}$ of size m x n bytes. 2 x 2 image block called mask is chosen from the source image matrix in row major order and transform it into frequency domain using the equation 1. Three bits of authenticating message/image are inserted at any three positions among 1st to 5th positions from LSB in each real part of each frequency component of source image block excluding the first frequency component of each image block. First component is used to maintain the imperceptibility and robustness. The insertion positions within each real frequency component are selected at even and odd positions from LSB in alternate bytes. After embedding the authenticating data in frequency domain then the IDFT is applied using equation 2 to transform from frequency to spatial domain. Then each time re-adjusting phase is applied to overcome the negativity and fractional value in spatial domain. Finally a control technique is used to reduce the noise. In this technique the unchanged intermediate positions and just after the maximum embedding position are consider here and adjust them in such a manner that the changes remain optimal before and after embedding. The reverse operation is performed at the receiving end to extract bits of authenticating message/image and message digest MD for authentication at destination.

In the frequency-domain all spatial-domain values are in form $a + i*b$, i.e. the complex frequency component. In IATFDDFT we cleverly chose the image block of size 2 x 2 from the source image to avoid the non-zero imaginary frequency component in the transformed value. The DFT for the 2 x 2 mask is $F(u, v) = \frac{1}{2}\sum\sum f(x, y)[\cos 2\Pi(ux/2 + vy/2) - i \sin 2\Pi(ux/2 + vy/2)] = \sum\sum f(x, y)[\cos\Pi(ux + vy) - i \sin\Pi(ux + vy)]$ where value of spatial variables x, y are 0, 1 and the value of frequency variables u, v are 0,1. For any values of x, y, u, and v the value of the imaginary components are zero and values of real components are either +1 or -1. So for transformation of all elements of 2 x 2 matrix will be in the form of $a + i*0$ i.e. either +a or -a. The proposed IATFDDFT technique embeds authenticating data into the frequency component of source image for any changes of frequency component it can affect the spectrum value which may change the quantum value in spatial domain. To maintain the balance in each mask first frequency component is used as re-adjust phase and remaining three of each mask is used to embed authenticating data.

In the proposed algorithm after embedding we have used inverse discrete Fourier transform (IDFT) to get the embedded image in spatial domain. Applying IDFT on identical mask with embedded data the quantum values may changes it can generate the following situation:

A. The converted value may by negative (-ve).

B. The converted value in spatial domain may be a number with non zero fractional value i.e. pure non integer number.

C. The converted value of each image byte may be greater the maximum value (i.e. 255).

The concept of re-adjust phase is to handle the above three serious problem by using the first frequency component of each mask. In this phase if the converted value is -ve or with fractional value then add 1 with the first frequency component in the mask and then apply IDFT. This repeating process continue until all are not will be non negative and non fractional.

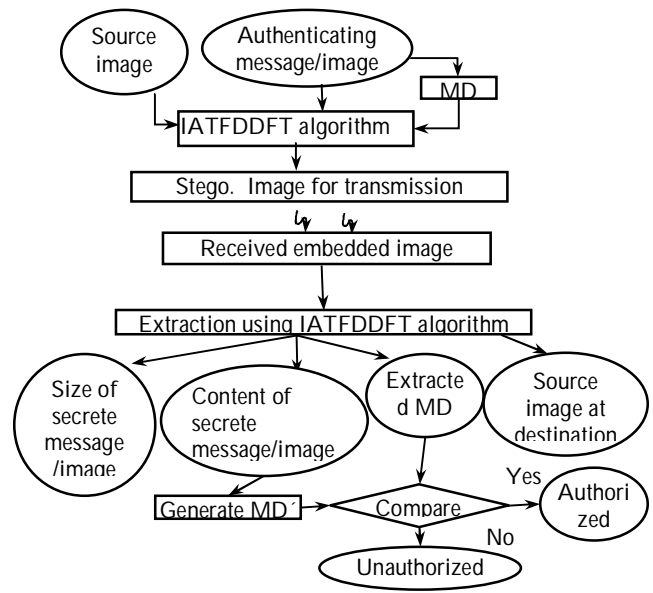

Fig. 1: Schematic diagram of IATFDDFT technique

Department of Computer Science, The University of Burdwan, Burdwan, West Bengal, India

For case C if the number is greater than the maximum value then subtract 2 from the first frequency component and then apply IDFT. This process is continuing until any value of the mask is greater than 255. The entire process of the IATFDDFT technique is given in Figure 1.

*A. Algorithm for Insertion*

In this algorithm all insertion is made in frequency domain i.e. each byte of source image in each mask of size 2 x 2 is transformed to frequency domain using DFT by equation 1. The IATFDDFT scheme uses colour image as the input to be authenticated by text message/image. The authenticating message/image bits size is 3*(m x n) – (MD+L) where MD and L are the message digest and dimension of the authenticating image respectively for the source image size m x n bytes.

Steps:

1. Obtain 128 bits message digest MD from the authenticating message/image.

2. Obtain the size of the authenticating message/image (32 bits, 16 bits for width and 16 bits for height)

3. Read authenticating message/image data do

   • Read source image matrix of size 2 × 2 mask from image matrix in row major order and apply DFT.

   • Extract authenticating message/image bit one by one.

   • Compute the position within the real frequency component where authenticating message/image bit is to be inserted (excluding 1st component).

   • Replace the authenticating message/image bit in the computed position within each real part.

4. Apply inverse DFT using identical mask.

5. Apply re-adjust phase if needed.

6. Apply control phase.

7. Repeat step 3 to step 6 for the whole authenticating message/image size, content and for message digest MD.

8. Stop.

*B. Algorithm for Extraction*

The authenticated image is received in spatial domain. During decoding the embedded image has been taken as the input and the authenticating message/image size, image content and message digest MD are extracted data from it. All extraction is done in frequency domain from frequency component.

Steps:

1. Read embedded source image matrix of size 2 × 2 mask from image matrix in row major order and apply DFT.

2. For each mask do

   • Compute the position within the real frequency part (excluding 1st frequency component of each mask) for each embedded image quantum value where authenticating message/ image bits are available.

   • Extract the message/image bit.

   • For each 8 (eight) bits extraction construct one alphabet/one primary (R/G/B) colour image.

3. Repeat step 1 and step 2 to complete decoding as per size of the authenticating message/image.

4. Obtain 128 bits message digest MD′ from the extracted authenticating message/image. Compare MD′ with extracted MD. If both are same the image is authorized else unauthorized.

5. Apply inverse DFT using identical mask.

6. Stop

*C. Example*

In this section the process of proposed IATFDDFT technique is figuratively presented sequentially. Consider the message string 'SACHIN' (Fig. 2a) to be embedded into the source image matrix as given in Fig. 2b. Fig. 2c shows the scheme for transformation of one 2 × 2 submatrix from spatial domain to frequency domain using DFT by equation 1. Here image bits are replaced by message bits at any three positions among 4

Department of Computer Science, The University of Burdwan, Burdwan, West Bengal, India

positions of real part (transformed value) of source transformed value from LSB. Inverse transformation IDFT of the embedded image is shown in Fig. 2d for transformation from frequency domain to spatial domain.

| Character | ASCII Code |
|-----------|------------|
| S | 01010011 |
| A | 01000001 |
| C | 01000011 |
| H | 01001000 |
| I | 01001000 |
| N | 01001110 |

Fig. 2a. Secrete Data

| 15 | 36 | 19 | 45 |
|----|----|----|----|
| 17 | 20 | 55 | 78 |
| 11 | 10 | 16 | 80 |
| 4  | 6  | 18 | 91 |
| 0  | 34 | 15 | 54 |
| 30 | 15 | 12 | 70 |

Fig. 2b. Source Image

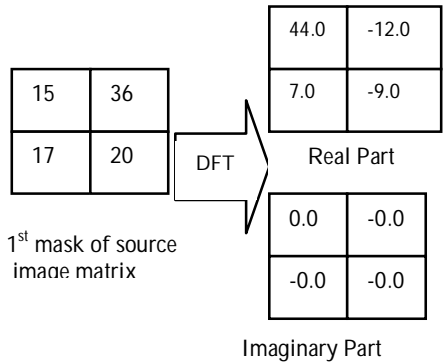

Fig.2c. Conversion of image matrix into frequency domain using DFT

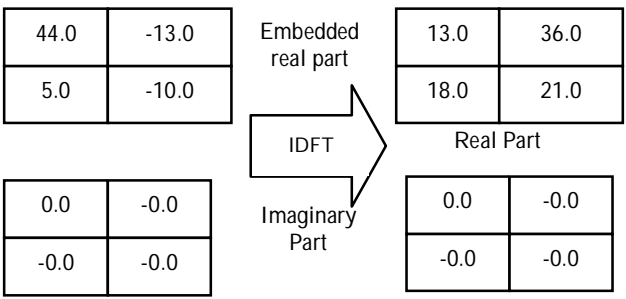

Fig. 2d: Conversion from frequency domain to spatial domain

### III. RESULT, COMPARISON and ANALYSIS

This section represents the results, discussion and a comparative study of the proposed technique IATFDDFT with the DCT-based watermarking method and QFT based watermarking method in terms of visual interpretation, image fidelity (IF), and peak signal-to noise ratio (PSNR) analysis and mean square error (MSE). In order to test the robustness of the scheme IATFDDFT, the technique is applied on more than 50 PPM gray images from which it may be reveille that the algorithm may overcome any type of attack like visual attack and statistical attack. The distinguishing of source and embedded image from human visual system is quite difficult. In this section some statistical and mathematical analysis is given. The original source images 'Baboon', 'Sailboat' are shown in Fig. 3a and Fig. 3b and 73728 bytes of information are embedded into each image. The dimension of each source gray images is 512 x 512 and the dimension of authenticating gray image is 270 x 270, shown in Fig. 3c. Fig. 3d and Fig. 3e are embedded images using IATFDDFT. Three bits of authenticating information are embedded at any three positions between 1st to 4th positions of real part of the frequency component excluding first component in each mask.

We use the peak-to-signal noise ratio (PSNR) to evaluate qualities of the stegoimages. Table I shows the noticeable amount of secrete data embedding is done with higher PSNR values for different source images. Table II shows the PSNR values for Lenna image in existing methods [22] like SCDFT, QFT and DCT. In all the techniques the dimension of Lenna JPEG image is 512 x 512. In all the existing technique the PSNRs are low, means bit-error rate are high but in the proposed scheme more bytes of authenticating data can be embedded and the PSNR values are significantly high, means bit-error rate is low. In DCT based watermarking scheme do not embed watermarks in every single block of image. Here selectively pick the regions that do not generate visible distortion for embedding, thus decreasing the authenticating data size. In QFT based watermarking compensation mark allows the watermark to be undetected even if the strength of it is high. For low compression factor it can not completely recover the embedded message. In IATFDDFT the average embedding capacity is 73728 bytes with higher average PSNR values 37.54 and completely recoverable the authenticating message/image.

## IV. CONCLUSION

IATFDDFT technique is an image authentication process in frequency domain to enhance the security compared to the existing algorithms. In compare to DCT and QFT based watermarking technique our algorithm is applicable for any type of color images authentication and strength is high. First frequency component in each mask is used for re-adjusting. The control technique is applied to optimized the noise addition as a result PSNR is increased with low MSE and IF is nearer to 1. In the proposed IATFDDFT authentication is done in frequency domain without changing visual property of the authenticated image. In IATFDDFT distortion of image and change of fidelity (like sharpness, brightness etc) is negligible.

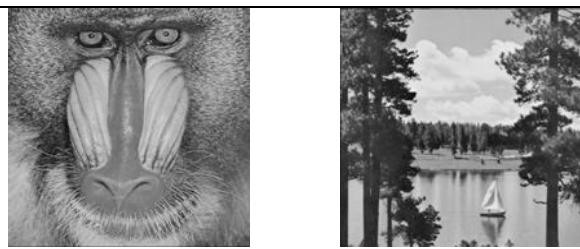

Fig. 3a. Source image 'Baboon'    Fig. 3b. Source image 'Sailboat'

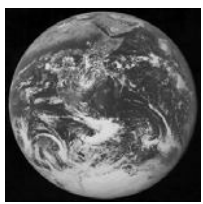

Fig. 3c. Authenticating image 'Earth'

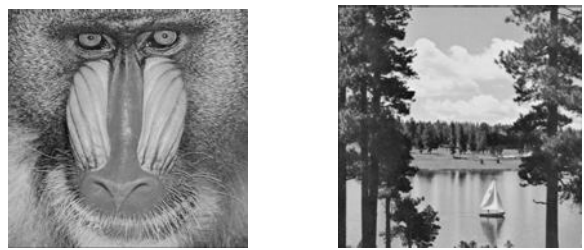

Fig. 3d. Embedded image using IATFDDFT    Fig. 3e. Embedded image using IATFDDFT

Fig. 3 : Comparison of fidelity in embedded 'Baboon' and 'Sailboat' image using IATFDDFT

Table I.

Capacities and PSNR, IF, and MSE in IATFDDFT

| Source images | Capacity (byte) | PSNR | IF | MSE |
|---|---|---|---|---|
| Sandiego | 73728 | 37.61 | .999594 | 11.277438 |
| Sailboat | 73728 | 37.31 | .999392 | 12.069550 |
| Woodland | 73728 | 37.47 | .999513 | 11.647601 |
| Baboon | 73728 | 37.75 | .999430 | 10.912079 |
| Average | 73728 | 37.54 | .999482 | 11.476667 |

Table II.

Capacities and PSNR for Lena image in the existing technique [22]

| Technique | Capacity(bytes) | PSNR |
|---|---|---|
| SCDFT | 3840 | 30.1024 |
| QFT | 3840 | 30.9283 |
| DCT | 3840 | 30.4046 |